\documentclass[]{aa}
\usepackage{graphicx,psfrag,amsmath}
\begin{document}

        \title{Vertical     distribution      of     Galactic     disk
        stars:
\thanks{Based  on  observations  made  at the  Observatoire  de  Haute
        Provence (France). Based on data from the Hipparcos astrometry
        satellite.}   
}
       \subtitle{III.  The Galactic disk surface mass density from red clump giants}
       \titlerunning{ The mass density in the Galactic plane}

   \author{O.     Bienaym\'e\inst{1},   C.    Soubiran\inst{2},   T.V.
    Mishenina\inst{3},  V.V.   Kovtyukh\inst{3},  A.  Siebert\inst{4}}
    \authorrunning{O. Bienaym\'e et al.}
   \mail{bienayme@astro.u-strasbg.fr}
   \institute{Observatoire  astronomique   de  Strasbourg,  UMR  7550,
              Universit\'e  Louis  Pasteur,  Strasbourg  France  
\and
              Observatoire  aquitain des  sciences  de l'univers,  UMR
              5804, BP 89, 33270 Floirac, France 
\and
Astronomical Observatory of Odessa
National University and Isaac Newton Institute of Chile,
Shevchenko Park, 65014, Odessa, Ukraine
\and
              Department of Astronomy/Steward Observatory, Tucson AZ 
}    
%
%
\date{Received 31 May 2005~/~Accepted 15 September 2005}
   \abstract{  We used  red clump  stars to  measure the  surface mass
density  of  the  Galactic  disk  in the  solar  neighbourhood.   High
resolution  spectra of  red  clump  stars towards  the  NGP have  been
obtained  with the  ELODIE spectrograph  at OHP  for  Tycho-2 selected
stars,  and  nearby Hipparcos  counterparts  were  also observed.   We
determined their  distances, velocities, and  metallicities to measure
the gravitational  force law perpendicular to the  Galactic plane.  As
in  most previous  studies,  we applied  one-parameter  models of  the
vertical  gravitational potential.   We obtained  a disk  surface mass
density   within  1.1\,kpc   of  the   Galactic   plane,  $\Sigma_{\rm
1.1\,kpc}\,=\,64\pm5\,\mathrm{M}_{\sun}   \mathrm{pc}^{-2}$,  with  an
excellent formal  accuracy, however  we found that  such one-parameter
models   can   underestimate   the   real   uncertainties.    Applying
two-parameter models, we derived more realistic estimates of the total
surface  mass   density  within  800\,pc  from   the  Galactic  plane,
$\Sigma_{\rm                   0.8\,kpc}$\,=\,57-66$\,\mathrm{M}_{\sun}
\mathrm{pc}^{-2}$,      and     within      1.1\,kpc,     $\Sigma_{\rm
1.1\,kpc}$\,=\,57-79$\,\mathrm{M}_{\sun}  \mathrm{pc}^{-2}$.  This can
be compared to  literature estimates of $\sim$40\,$\,\mathrm{M}_{\sun}
\mathrm{pc}^{-2}$    in   stars   and    to   13\,$\,\mathrm{M}_{\sun}
\mathrm{pc}^{-2}$  in the less  accurately measured  ISM contribution.
We conclude that there is no  evidence of large amounts of dark matter
in the  disk and, furthermore, that  the dark matter halo  is round or
not vey much flattened.

  A by-product of  this study is the determination  of the half period
of oscillation  by the Sun through the  Galactic plane, $42\pm2\,$Myr,
which cannot  be related to  the possible period of  large terrestrial
impact craters $\sim$ 33-37\,Myr.
}
   \maketitle
%
%

\section{Introduction}

This  paper  is the  extension  of  the  previous works  published  by
Soubiran et al.  (\cite{sou03},  hereafter Paper~I) and Siebert et al.
(\cite{sie03}, hereafter Paper~II), which probed the properties of red
clump stars within 100\,pc of  the Sun and at larger distances towards
the North Galactic Pole (NGP).  We obtained a new determination of the
local surface mass  density.  We discuss this new  result with respect
to  recent   works,  and  we   comment  on  the  different   ways  the
uncertainties have  been estimated and the  resulting consequences for
the estimated mass density of the Galactic disk.

Since the pioneering works  of Kapteyn (\cite{kap22}) and, later, Oort
(\cite{oor32,oor60}),  regular  improvements  have  been  obtained  in
determining the  vertical Galactic potential, thereby  allowing one to
constrain the  determination of the  total volume mass density  in the
solar neighbourhood,  now called  the Oort limit.  Two decades  ago, a
seminal improvement was achieved  by Bahcall (\cite{bah84}), who built
a consistent  Galactic vertical potential linked  to current knowledge
of   the  kinematics   and  the   density  distributions   of  stellar
populations.   Bienaym\'e  et  al.   (\cite{bie87})  followed  a  very
similar  approach  by  constraining  the Galactic  vertical  potential
through global Galactic  star counts and the current  knowledge on the
stellar population kinematics.

Later,  a  major   step  forward  was  made  by   Kuijken  \&  Gilmore
(\cite{kg89}) with a new sample of K dwarfs towards the South Galactic
Pole (SGP) tracing the vertical potential. They used the same stars to
measure both  the vertical density distribution  and kinematics.  This
had the  immediate consequence of  considerably reducing uncertainties
existing  in previous  works,  where different  samples  were used  to
determine both the vertical density and kinematics.

Thereafter,  regular  advances  occured  with  improved  stellar
samples and  accurate corrections of  systematic effects by  Flynn and
collaborators (\cite{fly94,hog98, hol00, hol04}),  in our Papers I and
II, and by other authors (for instance Korchagin et al., \cite{kor03}).

A  decisive   moment  was,  of   course,  the  arrival   of  Hipparcos
observations (\cite{esa97})  that allows precise calibration\footnote{
We note  that the  correction of systematic  effects should  be easier
with Hipparcos data, since the distribution of errors is understood so
well. However, the way the correction of systematic effects is applied
always  depends on the  astrophysical question  examined. The  bias of
Lutz-Kelker,  Malmquist or  others  must be  cautiously considered  to
achieve a proper  correction; see for instance a  discussion by Arenou
et  al.   (\cite{are99}).  }    of  stellar  parallaxes  and  absolute
magnitudes.  This has allowed  robust estimation of distances and also
useful measurement of tangential velocities.  An immediate application
has consisted  in probing  the potential close  to the  Galactic plane
within 100-200\,pc  directly. Likewise, Hipparcos  data gave immediate
access  to  the  Oort  limit  (Pham  \cite{pha98},  Cr\'ez\'e  et  al.
\cite{cre98a,cre98b},  Holmberg \&  Flynn  \cite{hol00}, Korchagin  et
al. \cite{kor03}).

In  this  general  context,  our paper  describes  the  observational
extension of the  red clump samples analysed in  Paper~II, and gives a
new dynamical  determination of the total local  surface mass density.
We  also  discuss  what  are  probably the  real  current  constraints
obtained on the surface mass  density, and comment on other results
obtained in previous papers.

There  is  no  perfect   agreement  yet  between  the  various  recent
determinations of the vertical potential perpendicular to the Galactic
plane.  Samples  remain small; methods  and analysis are  probably not
yet  optimized;  and some  assumptions,  like  full  phase mixing  and
stationarity,  are  difficult   to  check.   Furthermore,  useful  and
complementary information  are not used  optimally like the  change of
metallicities with kinematics.

Even if many systematic effects can now be conveniently considered and
corrected,  the lack  of large  unbiased stellar  samples  with radial
velocities  prevent  us from  examining  the  stationarity of  stellar
tracers in detail, which is  a central question that will need further
examination. We  may, however, note that the  Hipparcos proper motions
and tangential velocities have allowed the 3D velocity distribution in
the solar neighbourhood to be  probed.  The phase mixing appears to be
\textquoteleft slow' for horizontal motions within the Galactic plane.
The corresponding period  of the epicyclic motions is  169\,Myr, and a
few streams are  still visible in the ($u,v$)  velocity space (Chereul
et al.   \cite{che99}, and Famaey et al.  \cite{fam04}).  For vertical
motions,  the oscillation  period  is shorter,  86\,Myr,  or half  the
epicyclic  period,  and only  one  velocity  stream  is still  clearly
visible (associated  with the  Hyades cluster); otherwise  the ($z,w$)
phase  space  corresponding  to  the  vertical  motions  seems  to  be
phase-mixed.

Future advances are expected with  the measurement of the disk surface
mass density towards regions  away from the solar neighbourhood. Local
kinematics still carries non-local information about the structure of
the disk: for instance, the  coupling between the vertical and ($u,v$)
horizontal  motions of stars  in the  solar neighbourhood  is directly
linked to the scale length of  the total mass distribution in the disk
(Bienaym\'e, \cite{bie99}).

Analysis of stellar  IR surveys, 2MASS, DENIS or  DIRBE (Smith et al.,
\cite{smi04}) will  allow minimization  the effects of  the extinction
and uncertainties  on distance scales.  Surveys like  the RAVE project
(Steinmetz, \cite{ste03}) will increase the number of available radial
velocities by one  or two orders of magnitude:  a few hundred thousand
bright  stars with  $I \le  12$.  The  next step,  an  ESA cornerstone
mission, will be the GAIA project (Perryman et al., \cite{per01}), but
new methods of analysis  should be prepared.  Classical analyses, like
the one  applied in  this paper, would  certainly be  insufficient for
fully investigating the huge amount of data expected.

In Sect.~2 we describe selection of the 3 samples that we use: a local
one and 2  distant ones towards the North  Galactic Pole, optimized to
include a  large fraction  of clump giants.   Section 3 is  devoted to
methods and explains how we determinated of the vertical potential and
the  disk surface mass  density.  Our  discussion and  conclusions are
given  in  Sects.   4  and  5.  In  Soubiran  et  al.   (\cite{sou05},
hereafter  Paper~IV), we describe  the improvement  of our  local and
distant  samples in  detail as  compared to  Papers~I and  II,  and we
analyse these  samples in  terms of the  properties of thin  and thick
disk populations.

\section{The survey}

To determine  the vertical force perpendicular to  the Galactic plane,
we measured the  spatial and the vertical velocity  distributions of a
test stellar population  as a function of vertical  height.  As far as
possible, this  test population must be homogeneous  and unbiased with
selection criteria that are  independent of velocity and distance.  It
must also  be in a  stationary state.  For  this purpose, we  used one
local and two distant samples  of red clump giants selected within the
same $B-V$ colour  window and the same absolute  magnitude window; our
selected NGP clump stars are  the distant counterparts of our selected
Hipparcos stars.  We  find that at magnitude $V=9.5$  towards the NGP,
half of  stars, with $0.9  \le B-V \le  1.1$ are clump stars  while at
magnitudes $V \le  7.5$, more than 80 percent  are clump stars. Redder
and bluer clump  stars do exist.  The blue cut  removes the most metal
poor  stars ([Fe/H]  $\le -0.60$)  very efficiently  but allows  us to
reach  fainter  magnitudes with  low  contamination  by main  sequence
stars.   The red  cut  was  applied to  minimize  the contribution  of
subgiant stars and of other giants  on their first ascent of the giant
branch.

The  distant sample  is the  extension  to larger  distances from  the
Galactic  plane of  the NGP  sample  that was  previously analysed  in
Papers~I and  II. It  was built  from a preliminary  list of  red clump
candidates   from  the   Tycho-2   star  catalogue   (H{\o}g  et   al.
\cite{hog00}).  High  resolution spectroscopic observations  were used
to confirm the red clump stars, to separate them from the other stars,
and to measure radial velocities. We also improved the local sample of
203 red  giants by measuring  new radial velocities  and metallicities
for 88 of  these stars.  The selection, observation,  and reduction of
the two samples is briefly described below and explained in Paper~IV.

\subsection{The Hipparcos red clump stars}

The local sample of 203 nearby red clump giants was  selected from
the Hipparcos catalogue according to the following criteria :
$$ \pi \ge 10\,\rm{mas}$$
$$ \delta_{\rm{ ICRS}} \ge -20\deg$$
$$0.9 \le B-V \le 1.1$$  
$$ 0  \le M_{\rm V} \le 1.3  $$ 
where     $\pi$     is     the     Hipparcos    parallax     and     $
\delta_{\rm{ICRS}}$\footnote{The   Hipparcos    star   positions   are
expressed  in  the   International  Celestial  Reference  System  (see
http://www.iers.org/iers/earth/icrs/icrs.html).}    the   declination.
The Johnson $B-V$ colour was  transformed from the Tycho-2 $B_{\rm
T}-V_{\rm T}$ colour by applying Eq.  1.3.20 from \cite{esa97}:
$$B-V  =  0.850  \,(B{\rm  _T}-V{\rm  _T}).$$  
The absolute magnitude $M_{\rm{V}}$  was computed with the $V$ magnitude
resulting from  the transformation of the  Hipparcos magnitude $H_{\rm
p}$ to  the Johnson  system with the  equation calibrated  by Harmanec
(\cite{har98}).

We   searched for  radial velocities  and metallicities  for these
stars in the literature.  Our source of spectroscopic metallicities is
the [Fe/H]  catalogue (Cayrel et al., \cite{cayrelstr01}),  in which we
found a fraction of our stars in common with McWilliam (\cite{mcW90})
and  Zhao et  al.  (\cite{zhaoet01}).   Unfortunately  metallicities by
Zhao et  al. (\cite{zhaoet01}) could  not be considered because  of an
error   in   their  temperature   scale,   and   those  by   McWilliam
(\cite{mcW90})  had  to  be  corrected  from  a  systematic  trend  as
described in Paper~IV.  Complementary data were obtained for 88 stars
observed  with  the  echelle  spectrograph ELODIE  in  February  2003,
October 2003, and  February 2004 at the Observatoire  de Haute Provence
(France) with signal to noise ratios  at 550 nm (S/N) ranging from 150
to  200.  We  measured  and determined  their  radial velocities,
[Fe/H]  metallicities,  and abundances  of  several chemical  elements.
Metallicities are missing for 7 remaining stars, among which there are
three binaries.

 A detailed  description of  the atmospheric parameters  and abundance
determination  is  given in  Kovtyukh  et  al.  (\cite{Kovtet05})  and
Mishenina   et  al.    (\cite{Mishet05}).    Briefly,  the   effective
temperatures  were  determined  with  line-depth ratios,  a  technique
similar  to the  one  developed by  Gray  (\cite{gray94}), leading  to
excellent precision  of 10-20~K.  The surface gravities  $\log g$ were
determined using  two different methods :  ionisation-balance for iron
and  fitting the wings  of a  Ca I  line profile.   For the  method of
ionisation-balance, we selected about 100  Fe I and 10 Fe II unblended
lines based  on the synthetic  spectra calculations obtained  with the
software STARSP  (Tsymbal, \cite{tsymbal96}). For  the profile-fitting
method, the Ca I line at 6162\,\AA, which is carefully investigated in
Cayrel  et al.   (\cite{CFFST96}), was  used.  The  gravities obtained
with these two methods show very good agreement, as shown in Mishenina
et al.  (\cite{Mishet05}).  The [Fe/H] determination is constrained by
the large  number of lines of  neutral iron present in  the spectra of
giants.  The iron abundances were determined from the equivalent width
of lines by applying the program of Kurucz WIDTH9.  The measurement of
the equivalent width of lines  was carried out with the program DECH20
(Galazutdinov\cite{gal94}).

\subsection{NGP K giants}

The  distant K  giant  sample  was drawn  from  the Tycho-2  catalogue
(H{\o}g  et al.,  \cite{hog00}).  We  applied similar  criteria  as in
Paper~I to build the list  of red clump candidates, just extending the
limiting  apparent  magnitudes  to  fainter  stars.   In  summary,  we
selected  stars in  two  fields close  to  the NGP.   The first  field
(radius $10\degr$,  hereafter F10) is  centred on the NGP,  the second
one  (radius $15\degr$,  hereafter  F15) is  centred  on the  Galactic
direction ($l=35.5\degr,b=+80\degr$),  avoiding the Coma  open cluster
area  (a  circular  field  of $4.5\degr$  radius  around  $l=221\degr,
b=84\degr$).  The total area effectively covered by our samples is 720
square degrees.  We selected stars with $0.9\le B-V\le1.1$, in the $V$
magnitude  range  7.0-10.6  for  F10,  and 7.0-9.5  for  F15  (Johnson
magnitudes transformed from Tycho2 ones).  Known Hipparcos dwarfs were
rejected.

A  total of 536  spectra were  obtained  with the  ELODIE echelle
spectrograph at  the Observatoire de Haute  Provence, corresponding to
523  different stars:  347 in  F10 and  176  in F15.   The spectra  have a
median S/N  ratio of  22 at  550\,nm.  This low  S/N is  sufficient to
estimate with   good accuracy the stellar  parameters, $T_{\rm eff}$,
gravity,  and [Fe/H] metallicity,  and absolute  magnitude $M_{\rm{V}}$
with the {\sc tgmet} method  (Katz et al.  \cite{kat98}), as previously
described in Paper~I.  {\sc tgmet} relies  on a  comparison by minimum
distance of the  target spectra to a library of  stars with well-known
parameters, also observed with  ELODIE (Soubiran et al.  \cite{sou98},
Prugniel \& Soubiran \cite{pru01}).  Since our previous study of clump
giants  at  the  NGP,  the  {\sc  tgmet}  library  has  been  improved
considerably.  Many stars  with well-determined atmospheric parameters
and with accurate Hipparcos parallaxes  have been added to the library
as   reference   stars  for   ($T_{\rm   eff}$,   $\log  g$,   [Fe/H],
$M_{\rm{V}}$).   The improvment  of the  {\sc tgmet}  library  and the
extended sample  is fully described  in Paper~IV.  Here we  just give
useful characteristics of the extended sample.

The  accuracy of  the {\sc  tgmet} results  was  assessed  with a
bootstrap  test on  reference stars  with very  reliable atmospheric
parameters and  absolute magnitudes.  A  rms scatter of 0.27  was 
obtained on  $M_{\rm{V}}$ and 0.13  on [Fe/H].  These values  give the
typical  accuracy  of  the   {\sc  tgmet}  results.   The  scatter  on
$M_{\rm{V}}$ corresponds to a 13\% error in distance.  As explained in
Paper~I, the absolute magnitudes from Hipparcos parallaxes of the {\sc
tgmet}  reference stars are  affected by  the Lutz-Kelker  bias.  This
causes an additional external error  that must be taken into account.
In Paper~II, the Lutz-Kelker bias  for local clump giants is estimated
to  be $-0.09$  mag corresponding  to a  systematic  overestimation of
distances by 4\%. We did not attempt to correct individual absolute
magnitudes  of the  reference stars,  because the  Lutz-Kelker  bias was
estimated from the luminosity  function of the parent population, which
is unknown for most giants of  the TGMET library, except for the clump
ones.

%
\begin{figure}[t]
\center
\includegraphics[width=6cm]{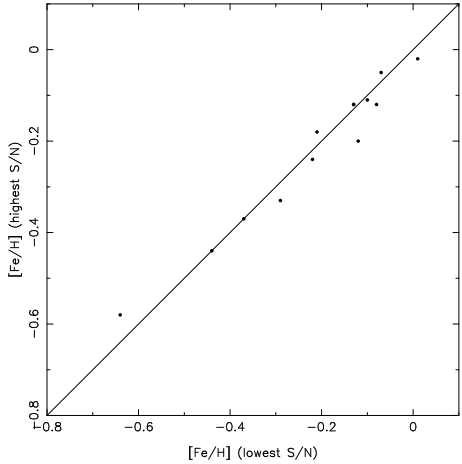}
\caption{Comparison of the {\sc tgmet} metallicities obtained for the 13
target stars observed twice.}
\label{f:internal_FeH}
\end{figure}

In order to  test the internal precision of {\sc  tgmet} on [Fe/H], we
 compared  the results obtained  for the 13 clump  giants observed
twice (Fig.~\ref{f:internal_FeH}).   As can be seen,  the agreement is
excellent ($\sigma=0.03$ dex). For $M_{\rm{V}}$, the  scatter is only
0.08 mag.

Once  the  stellar  parameters   ($T_{\rm  eff}$,  $\log  g$,  [Fe/H],
$M_{\rm{V}}$) were determined for each  of the 523 target stars of the
NGP  sample, $M_{\rm{V}}$  was used  to  identify the  real red  clump
giants from the dwarfs and  subgiants and to compute their distances.
ELODIE radial  velocities, Tycho-2  proper motions, and  distances were
 combined  to compute 3D velocities ($U,V,W$) with respect
to the Sun.  According to typical errors of 0.1\,km.s$^{-1}$ in radial
velocities, 15\,\% in  distances, 1.4\,mas.yr$^{-1}$ in proper motions
at a mean distance of 470\,pc for F10, and 1.2 mas.yr$^{-1}$ in proper
motions at a mean distance of  335\,pc for F15, the mean errors on the
two velocity  components $U$ and  $V$ are 5.6 and  4.0\,km.s$^{-1}$ in
F10 et F15,  respectively, while it is 0.1\,km.s$^{-1}$  on $W$ in both
fields.

\begin{figure}[t]
\center
\includegraphics[width=6cm]{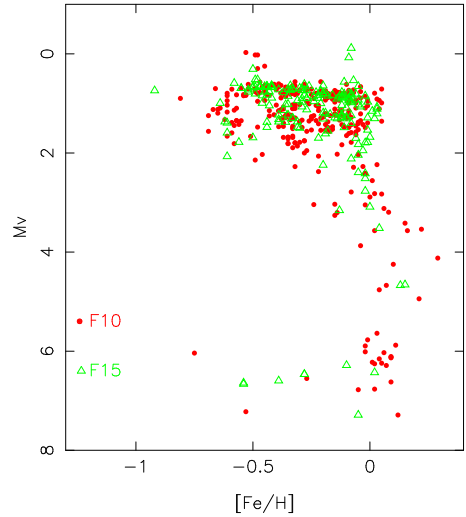}
\caption{Distribution  of the 523  target stars  in the  metallicity -
absolute magnitude plane.}
\label{f:pgn_fehMv}
\end{figure}

The 523 target stars are represented in Fig.~\ref{f:pgn_fehMv}, in the
plane metallicity - absolute magnitude, with different colours for F10
and F15.  The F10 field is more contaminated by dwarfs.

 \subsection{Binaries} Binarity  can be  recognized from the  shape of
the correlation function of the  spectra with a stellar template. None
of  the 536  target spectra  correspond to  a clear  SB2 (double-lined
spectroscopic binary),  but a dozen  of them present an  asymmetric or
enlarged profile  that could  be the signature  of a  companion.  Only
multi-epoch measurements could establish  the binarity fraction in our
sample precisely, but  it seems to be small as  also found by Setiawan
et  al. (\cite{set04}).   This was  expected,  as for  a short  period
binary, the  system would merge  during the giant phase.   Moreover, a
binary system with  a red clump giant and  a sufficiently bright star,
i.e.   spectral type  earlier than  K, is  expected to  have  a colour
outside our colour selection interval.   The only star that was really
a  problem  for {\sc  tgmet}  is  TYC\,1470-34-1,  because of  a  very
enlarged  profile,  which can  be  due either  to  a  companion or  to
rotation, so this star was removed from the following analysis.
 
\subsection{Kinematics and metallicity}
We note  a deficiency of  stars with metallicities [Fe/H]  $\le -0.6$,
although   this   is   not   specific   to  our   red   clump   sample
(Fig.~\ref{f:pgn_fehMv}).  Our selection criteria does not favour very
low  metallicity stars  for which  the  red clump  covers an  extended
colour interval, much larger than our colour interval selection.  This
lack of stars with [Fe/H] $\le -0.60$ exists also in distance-complete
samples of dwarfs,  for instance in Reddy et  al.  (\cite{red03}).  It
is  also  a  result  obtained  by  Haywood  (\cite{hay01,hay02})  when
revisiting the  metallicity distribution of  nearby G dwarf  stars.  A
more complete discussion of this  aspect of our sample is presented in
Paper~IV.

 For  our $K_z$ analysis,  we rejected  stars with  [Fe/H] metallicities
 outside      the      metallicity      interval      [$-0.25,+0.10$].
 Table~\ref{t:samples-caract} gives  the number of stars  used for the
 $K_z$ determination.

Forthy clump stars have  [Fe/H] abundances within $[-0.6,-0.45]$ and a
mean  vertical  velocity  of  $-20$\,km.s$^{-1}$,  which  differs  (at
3\,$\sigma$) from  the usual  or expected $-8$\,km.s$^{-1}$.   The 116
clump stars with [Fe/H]  abundances within $[-0.45,-0.25]$ have a high
velocity dispersion of $29.4$\,km.s$^{-1}$, and their vertical density
distibution  decreases slowly  over the  limited $z$-extension  of our
samples. Their  density decreases too gradually to  bring an efficient
constraint  on our present  $K_z$ determination.   Moreover, including
these  stars degrades the  analysis by  increasing the  uncertainty on
$K_z$  drastically.  For  all these  reasons, we  rejected  the lowest
metallicity stars from our analysis.

A small  fraction of stars have  [Fe/H] $\ge +0.10$.   These stars (24
stars  in the  local sample  and 1  in the  NGP cone  samples)  have a
relatively        low        vertical       velocity        dispersion
$\sigma_w$=10\,km.s$^{-1}$; they also have a correspondingly low scale
height.   In this  study, including  or rejecting  these stars  has no
influence: here they  do not constrain the total  surface mass density
of the disk, and have not been included in the analysis.

\begin{table}
\caption{Number  of  observed   stars  and number  of
identified red clump K-giants.} \center
\label{t:samples-caract}
\begin{tabular}{l c c c }\\
\hline
\hline
 & Hipparcos & Field 1 & Field2\\
\hline
Area (sq.deg.) & & 309 & 410 \\
$V_{\rm lim}$ & & 7-9.5 & 7-10.6\\
full sample & 203 & 176 & 347 \\
red clump $M_{\rm V}: 0\,{\rm to}\,1.3$ & 203 & 124 & 204  \\
Fe cuts    $-0.25$ to $+0.10$ & 152 & ~67 & 100 \\
\hline
\end{tabular}
\end{table}

\section{Volume and surface mass density determinations }

\subsection{The Oort limit}

Thanks to the  Hipparcos data, the Galactic potential  was  probed
in the  first two  hundred parsecs from  the Sun, giving   excellent
accuracy  for   determining   the  Oort limit,  i.e.  the  total
volume density in the Galactic plane at the solar position.

A first set of studies based on A to F Hipparcos stars within a sphere
of  about  125\,pc,  have   been  published  by  Pham  (\cite{pha98}),
Cr\'ez\'e  et al.   (\cite{cre98a,cre98b}), and  by Holmberg  \& Flynn
(\cite{hol00}).   Their  results  all  agree within  1-$\sigma$  error
limits, with  differences depending probably  on the various  ways the
potential  has been parameterized.   Holmberg \&  Flynn (\cite{hol00})
model  the  local  potential  according  to  a  set  of  disk  stellar
populations  and also  a  thin disk  of  gas, while  Cr\'ez\'e et  al.
(\cite{cre98a,cre98b}) assume  a simple quadratic  potential.  We note
that an  important limitation  has been the  lack of  published radial
velocities,  limiting  both  the  accuracy  of the  modelling  of  the
kinematics and the possibility of checking the stationary state of the
various  samples used  for these  analyses independently.   From these
studies,  we will  consider (see  Holmberg \&  Flynn  \cite{hol04} and
Paper~II)  that the  Oort  limit is  determined  as $\rho_{\rm  total}
(z=0)= 0.100 \pm  0.010\,\mathrm{M}_{\sun} \mathrm{pc}^{-3}$. The Oort
limit includes the local mass density from both the disk and dark halo
components.

Recently  Korchagin  et  al.   (\cite{kor03}), using  Hipparcos  data,
analyse  the vertical  potential  at slightly  larger distances.   The
tracer  stars are  giant stars  that are  brighter than  clump giants,
within a vertical cylinder of  200\,pc radius and an extension of $\pm
400$\,pc out of the Galactic plane.  For the dynamical estimate of the
local   volume   density,   they   obtain:   $\rho_{\rm   total}(z=0)=
0.100\pm0.005\,\mathrm{M}_{\sun}     \mathrm{pc}^{-3}$.     A    small
improvement could  perhaps still be achieved using  {\sc 2mass} colour
to  minimize  uncertainties  on  extinction, distances,  and  vertical
velocities.

\subsection{The total disk surface mass density.}

\subsection*{Model}
	To analyse the vertical  density and velocity distributions of
our samples  and to measure the  surface mass density  of the Galactic
disk,  we  model the  $f(z,w)$  distribution  function,  with $z$  the
vertical  position, $w$ the  vertical velocity  relative to  the Local
Standard  of Rest,  and we  adjusted the  free model  parameters  by a
least-square fitting to the  apparent magnitude star counts $a(m)$ and
to the vertical velocity dispersions in different magnitude intervals.
We simultaneously adjusted the model to the data from the 3 samples.

The  distribution function  is modeled  as the  sum of  two isothermal
components, according to
\begin{eqnarray}
\label{e:df}
f(z,  w)=   \sum_{k=1,2}
\frac{c_{k}}{\sqrt{2\pi}\sigma_{k}}\,  \exp^{-\left(\Phi(z)+\frac{1}{2}w^2  \right) /\sigma_{k}^2}.
\end{eqnarray}
For  the total  vertical potential  $\Phi(z)$, we used  the parametric
 expression proposed by Kuijken \& Gilmore (\cite{kg89}):
\begin{eqnarray}
\label{e:pot}
\Phi(z)\sim \Sigma_0 \left(\sqrt{z^2+D^2}-D\right) +\rho_{\rm eff}\, z^2 \, 
\end{eqnarray}
where the potential  is related, through the Poisson  equation, to the
vertical  distribution of  the  total volume  mass density  $\rho_{\rm
total}(z)$.  The $z$-integration gives  the total surface mass density
within $\pm\,z$ from the Galactic plane:

$$ \Sigma_{z\,{\rm kpc}}=\Sigma(<|z|)=\frac{\Sigma_0\, z}{\sqrt{z^2+D^2}} + 2 \rho_{\rm eff}\, z\,.$$

This parametric  law models the  vertical mass density by  two density
components.   One  component   mimics  the  locally  constant  density
$\rho_{\rm eff}$ of  a round or slightly flattened  halo.  It produces
 a  vertical quadratic  potential locally.  The other  component mimics
the  potential of  a  flat  disk; one  parameter,  $\Sigma_0$, is  its
surface mass density while the other, $D$, is its half thickness.

\subsection*{Data}
Compared  to  our previous  work  (Paper~II), we  have
increased  the number  of observed  stars  (726 against  387) and  the
limiting distances  by a  factor 1.26.  We  also obtained  new or
first [Fe/H]  determinations and built a complete  sample of Hipparcos
red clump stars with measured metallicities.  

Due to the lack of stars in the range 100-300\,pc, our samples are not
suited  to constrain the  potential in  this region  efficiently. More
generally,  pencil beam  samples towards  the Galactic  poles  are not
adequate   for  measuring   the  Oort   limit   without  supplementary
assumptions on  the shape of  the vertical potential.  Since  the Oort
limit has been  previously and accurately measured, we  will adopt for
the  volume  mass  density   the  value  discussed  above:  $\rho_{\rm
total}(z=0)\,=\,0.10\pm0.01\,\mathrm{M}_{\sun} \mathrm{pc}^{-3}$.

\subsection*{Fitting the vertical potential with {\it one} free parameter:}

\begin{figure}[hbtp]
\begin{center}
\includegraphics[width=9cm]{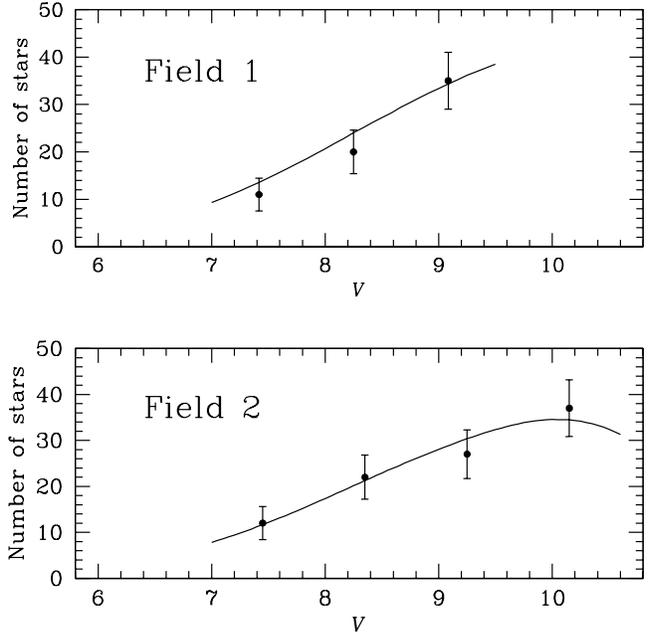}
\caption{ Observed  red clump  counts (circles), respectively  0.9 and
0.833 magnitude  wide bins for the  fields 1 and  2 towards the  NGP (local
sample and both NGP samples are fitted simultaneously). The continous line
indicates the best-fit model for the one-parameter potential.}

\label{f:am}
\end{center}
\end{figure}

\begin{figure}[hbtp]
\begin{center}
\includegraphics[width=9cm]{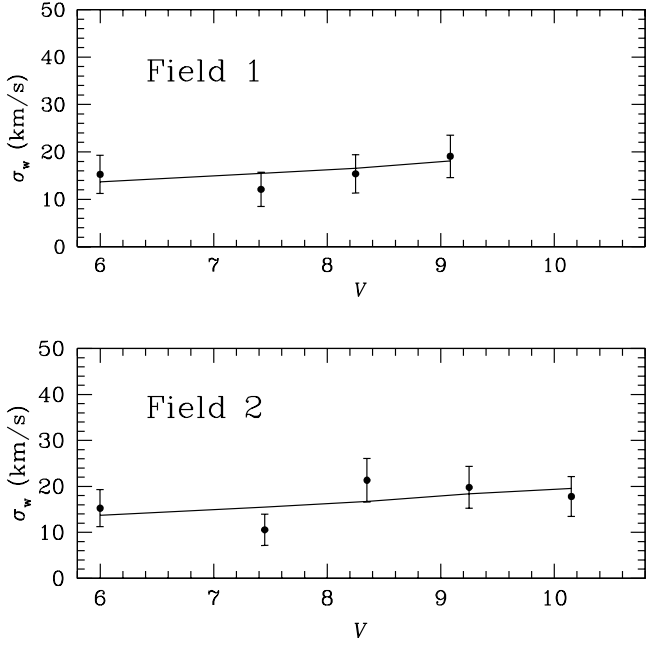}
\caption{Vertical   velocity    dispersion   versus   magnitude   (see
Fig.\,\ref{f:am}).   Velocity dispersions  of  Hipparcos stars  within
100\,pc are plotted at $V$\,=\,6.}
\label{f:sig}
\end{center}
\end{figure}

 A  least-square minimization is  applied to  the observed  star count
distributions  {\it and}  vertical  velocity dispersion  distributions
(see     continuous      lines     Fig.~\ref{f:am}-\ref{f:sig}     and
Table~\ref{t:fit-components}).   We  fit   the  data  from  the  three
samples, local and distant.  We  fit all the data {\it simultaneously}
to improve the parameter estimation and also the estimate of errors on
parameters.

Individual  errors on  distances  are  small for  stars  from the  NGP
samples,  about  13\%, but  this  does  not  contribute to  the  model
uncertainties since the analysed  quantities (star counts and vertical
velocities)  are independent of  distance. Vertical  velocities remain
slightly  affected by  uncertainties  on distances  through the  small
contribution  of the  projection  of proper  motions  on the  vertical
$z$-direction.   The uncertainty on  distances, however,  reflects the
accuracy  achieved  on absolute  magnitude  determination for  distant
stars  (about 0.27  mag) and  our  ability to  identify distant  clump
stars.  Clump  star absolute magnitudes are  normally distributed (see
Paper~II) around  $M_{\rm v}=0.74$ with  a dispersion of 0.25,  and we
select them in the range $0  \le M_{\rm v} \le 1.3$.  Our results also
depend on  the absolute magnitude  calibration of nearby  giant stars,
which is also explained in Paper~II.


\begin{table}
\caption{ Red clump disk properties.}
\center
\label{t:fit-components}
\begin{tabular}{l c }\\
\hline
\hline
[Fe/H] from $-0.25$ to +0.10\\
local relative density & velocity dispersion (km\,s$^{-1}$)\\
\hline
$c_{1}= 0.73\pm0.12$   & $\sigma_{1}=11.0\pm1.1$ \\
$c_{2}= 0.27\pm0.12$   & $\sigma_{2}=19.7\pm1.7$ \\
\hline
\end{tabular}
\end{table}        

We  may  consider  that  the  main  source  of  uncertainties  is  the
restricted size  of the samples.  They are  split into magnitude-bins,
and the uncertainty is dominated  by Poisson fluctuations. For a given
bin of magnitude, the error  on $a(m)$ is $\sqrt{a(m)}$, and the error
on $\sigma_w$  is $\sigma_w/\sqrt{2a(m)}$, where $a(m)$  is the number
of stars in the  bin.  Errors given in Table~\ref{t:fit-potential} and
elsewhere are deduced from the diagonal of the covariance matrix given
by the least-square fit.

\subsection*{Results}
We assume  that the Galactic  dark matter component needed  to explain
the  flat rotation  curve  of our  Galaxy  is spherical  and that  its
density     in    the     solar     neighbourhood    is     $\rho_{\rm
eff}=0.007\,\mathrm{M}_{\sun}  \mathrm{pc}^{-3}$  (Holmberg \&  Flynn,
\cite{hol00}).   Adopting  this   value,  along  with  the  previously
mentioned  local  density   $\rho_{\rm  total}(z=0)$,  then  only  one
parameter  of  our  potential   expression  is  free,  since  the  two
parameters $\Sigma_0$ and $D$ are related through
$$\rho_{\rm total}(z=0)=\rho_{\rm  eff}+\Sigma_0/(2\,D)\,.$$ 

We find  $D=260\pm24$\,pc, and the  surface density within  800\,pc is
found  to   be  $\Sigma_{800\,{\rm  pc}}\,=\,60\pm5\,\mathrm{M}_{\sun}
\mathrm{pc}^{-2}$.   Within  1.1\,kpc,  we  obtain  $\Sigma_{1.1\,{\rm
kpc}}\,=\,64\pm5\,\mathrm{M}_{\sun}   \mathrm{pc}^{-2}$.    The   most
recent                 determination,                 $\Sigma_{1.1{\rm
kpc}}$=\,71$\pm$6$\,\mathrm{M}_{\sun}  \mathrm{pc}^{-2}$,  obtained by
Holmberg \&  Flynn (\cite{hol04}) is  comparable. Their study  and ours
have many similarities, with giant stars in the same range of apparent
magnitudes,  accurate  absolute   magnitude  and  distance  estimates,
similar number  of stars and  apparently similar kinematics,  but also
many   differences:  NGP  versus   SGP  stars,   spectroscopic  versus
photometric distances,  clump giants versus  a sample with  a slightly
larger range in colours and absolute magnitudes.

With   an    estimated   mass   density    of   $53\,\mathrm{M}_{\sun}
\mathrm{pc}^{-2}$   for  the  visible   matter  (Holmberg   \&  Flynn,
\cite{hol04}) (see Sect. \ref{a:stellarSM}), there is no need {\it a
priori} to add matter to the model, for example, by flattening the dark
corona to explain our measured local surface density.

The other  parameters (relative densities and  velocity dispersions of
stellar components)  are explained in  Table~\ref{t:fit-components}: the
two fixed parameters are the  Sun's position above the Galactic plane,
$z_0$\,=\,15\,pc,  and the  Sun's  vertical velocity  $w_0$\,=8\,${\rm
km\,s}^{-1}$.

%
\begin{figure}[htbp]
\includegraphics[width=9cm,angle=0]{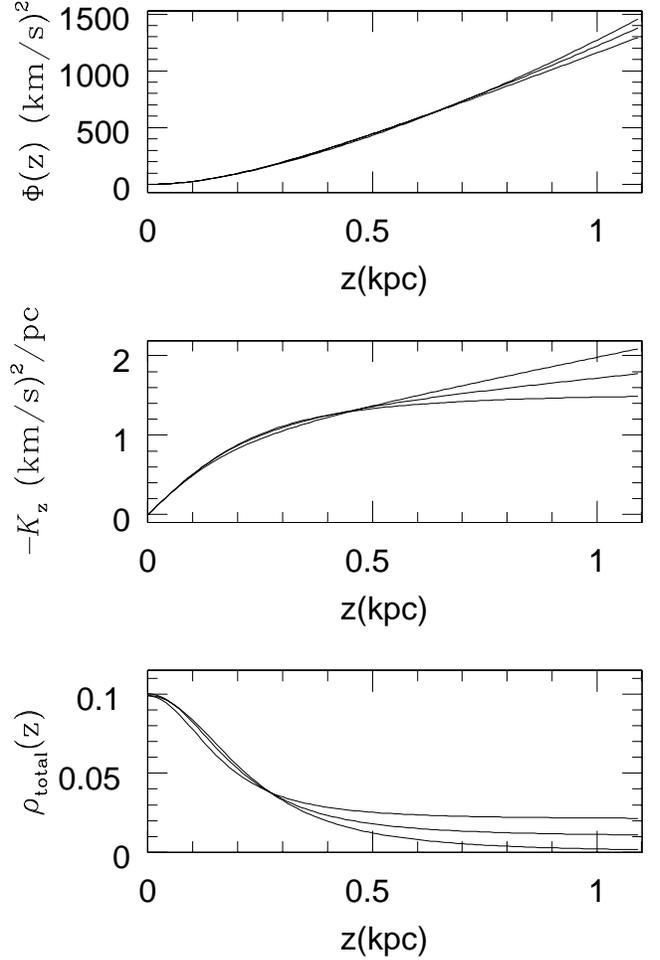}
\caption{The vertical potential (top),  the $K_{\rm z}$ force (middle),
and the total volume  density (bottom) for the three solutions ($\rho_{\rm
eff}=0,0.01,0.021\mathrm{M}_{\sun}    \mathrm{pc}^{-3}$)    given   in
Table\,\ref{t:fit-potential}.}
\label{f:Pot-Kz}
\end{figure}
%
\begin{table}
\caption{Solutions for the thickness of the vertical potential and the
total  surface  mass density.  Uncertainties  on  D  and $\Sigma$  are
$\sim\,9\%$} 
\center
\label{t:fit-potential}
\begin{tabular}{l  c c  c c c}\\ 
\hline  
\hline 
$\rho_{\rm  eff}$ &  $D$  & $\Sigma_{800\,{\rm pc}}$& $\Sigma_{1.1\,{\rm kpc}}$ \\ 
\hline 
$\mathrm{M}_{\sun}  \mathrm{pc}^{-3}$   &  pc   & $\mathrm{M}_{\sun} \mathrm{pc}^{-2}$  & $\mathrm{M}_{\sun} \mathrm{pc}^{-2}$ \\
\hline
0.00  & 287 & 57 & 57  \\ 
0.007 & 260 & 60 & 64  \\ 
0.01  & 249 & 61 & 67  \\ 
0.021 & 205 & 66 & 79  \\ 
\hline 
\end{tabular} 
\end{table} 
\subsection*{Fitting the vertical potential with {\it two} free parameters}

Adjusting the  model (Eq.\,\ref{e:pot})  with a single  free parameter
gives a  satisfying result with small  error bars (9  percent), but it
does  not give  information  on correlations  with  the other  (fixed)
parameters and does not tell us much about the range of possible other
solutions that are compatible with  the observations.  But it is known
that  in  practice,  the  change  of  $\rho_{\rm  eff}$  does  have  a
significant   impact  on  solutions:   its  correlation   with  others
parameters is discussed by Gould (\cite{gou90}).

Adjusting both $\Sigma_0$  and $\rho_{\rm eff}$ we find  that the best
fit  is obtained with  $\rho_{\rm eff}=0$,  while other  parameters are
very close to these given in Table~\ref{t:fit-components}.  
Solutions  within  1-$\sigma$  from  this  best  fit  give  the  range
0-0.021$\,\mathrm{M}_{\sun} \mathrm{pc}^{-2}$ for the $\rho_{\rm eff}$
parameter.

Table~\ref{t:fit-potential}  shows some  $\rho_{\rm  eff}$ values  and
solutions that  still result in  acceptable fits to  our observations.
Thus  acceptable  solutions  for  $\Sigma_{1.1\,{\rm  kpc}}$  cover  a
greater  range of  22 $\mathrm{M}_{\sun}  \mathrm{pc}^{-2}$  than does
using a  model with just  one free parameter.  This  is understandable
considering  that the best  determined quantity  is the  vertical {\it
potential} (the  adjusted quantity  in Eq.~\ref{e:df}), not  the $K_z$
{\it  force}.   The derivative  of  the  potential  gives the  surface
density,  which  is  proportional  to  the $K_z$  force,  and  similar
potentials may  produce different  $K_z$ forces and  surface densities
(see Fig.~\ref{f:Pot-Kz}).

This probably  explains the scatter  in determinations of  the surface
  density in previous works in the  80's and 90's.  
  These  correlations were already  discussed in  detail in  the same
  context by Gould (\cite{gou90})  to explain differences in the local
  surface mass density estimated by various authors.

 We  also  note   that  Eq.\,\ref{e:pot}  parameterizes  the  vertical
potential and  the total vertical mass distribution;  however, the two
apparent  components of the  r.h.s of  Eq.\,\ref{e:pot} should  not be
strictly read  as being  one component for  the stellar and  gas disks
(with a total  surface mass density $\Sigma_0$) and  another one for a
round or  flattened dark  matter halo.  Only  the total  vertical mass
density    distribution    (plotted    at    the   the    bottom    of
Fig.~\ref{f:Pot-Kz}) is constrained by  our data; from that figure, we
notice that  the three  plotted solutions $\rho_{total}(z)$  are quite
similar in the range of distances 0 to 400\,pc.

\section{Discussion}

An abundant literature and a  variety of results exist on the vertical
$K_z$ force and  the different methods applied to  constrain the total
local mass volume density, the  Oort limit.  The oldest papers must be
read with care, since systematic  bias on distances was more difficult
to check before the  Hipparcos satellite, but the techniques developed
and comments  remain valid.  Stellar  samples are extracted  from thin
and thick disk  populations and are represented as  the combination of
isothermal populations. Kuijken (\cite{kui91}) proposed
an  original  modeling  with  a  continuum  set  of  such  isothermal
populations.  We estimate that the most decisive aspect to understand
the  differences  between  authors  concerns  the  various  techniques
applied in modeling the vertical potential.

For instance, a powerful  technique is non-parametric modeling of the
$K_z$ force (Chen et al.   \cite{che03} and references in Philip \& Lu
\cite{phi89}).   These non-parametric  $K_z$ determinations  have been
achieved without smoothing or  regularisation (or, for instance, without
a positivity  condition on the  mass distribution), and  the resulting
published $K_z$  forces show large oscillations  that certainly result
from the small sample sizes.  One consequence is that these oscillations
have no physical interpretation; for instance, the resulting total mass
distribution is not positive everywhere. We expect that a conveniently
applied regularisation should be sufficient for making these methods more
reliable in this context.

On  the other  hand, the  parametric modeling  consists in  assuming a
global shape  for the vertical potential.   Bahcall (\cite{bah84}) and
recently Holmberg  \& Flynn (\cite{hol04})  have assumed a  total disk
mass  proportional to the  observed disks  of gas  and stars,  with an
extra component of dark  matter.  This last component, proportional to
one of  the known  components, is adjusted  to constrain  the vertical
potential.  Kuijken  \& Gilmore  (\cite{kg89}) also proposed  a simple
analytical model  (the model used in  this study) and  adjusted one of
the parameters.

The advantage  of using such  {\it a priori} knowledge  and realistic
models is to minimize the number  of free parameters and to reduce the
measured  uncertainties.  But  this decrease  does not  mean  that the
adjusted parameters  have been really obtained with   better accuracy,
since the information on correlations is lost.

For example,  in this paper,  we have adjusted the  vertical potential
with  two free  parameters; as  a consequence,  the formal  errors are
larger while the observational accuracy of our data is probably better
or, at  least of  similar quality, than  in the other  recent studies.
But this adjustment  with two free parameters allows  us to probe more
realistic and  general potentials.  After analysing our  data, we drew
the  following  conclusion:   the  uncertainty  on  $\Sigma_{1.1\,{\rm
kpc}}=\,68\pm11\,\mathrm{M}_{\sun}        \mathrm{pc}^{-2}$       (see
Table\,\ref{t:fit-potential}) is  still large, about  sixteen percent,
and this must also hold for previously published analyses.  \\

\subsection{The visible surface mass density:}
\label{a:stellarSM}

  Similar values  of $\Sigma_*$, the stellar surface  mass density, at
the solar position have  been proposed in recent works: $\Sigma_*\sim$
25$\,\mathrm{M}_{\sun} \mathrm{pc}^{-2}$ (Chabrier, \cite{cha01}), but
also  29\,$\mathrm{M}_{\sun}  \mathrm{pc}^{-2}$  (Holmberg  \&  Flynn,
\cite{hol00})  and 28$\,\mathrm{M}_{\sun}  \mathrm{pc}^{-2}$  from the
Besan\c  con  Galaxy model  (Robin,  \cite{rob03}).   The brown  dwarf
contributions  to  the  surface  mass  density  have  been  estimated:
6$\,\mathrm{M}_{\sun}    \mathrm{pc}^{-2}$    (Holmberg   \&    Flynn,
\cite{hol04})  or  3$\,\mathrm{M}_{\sun} \mathrm{pc}^{-2}$  (Chabrier,
\cite{cha02}).  More  difficult is  to estimate the  ISM contribution;
Holmberg  \&   Flynn  (\cite{hol00})  proposed  13$\,\mathrm{M}_{\sun}
\mathrm{pc}^{-2}$, but this quantity is uncertain. For instance it has
been proposed  that all  the dark matter  could contribute to  the ISM
disk  component.  Adding  all known  contributions, Holmberg  \& Flynn
(\cite{hol04})  propose 53$\,\mathrm{M}_{\sun}  \mathrm{pc}^{-2}$, the
value we adopt here.

\subsection{Are our $K_z$ solutions compatible  with our current 
knowledge of the Galactic rotation curve? }

 To   answer   this   question,    we   simplified   and   adopted   a
double-exponential  density   distribution  for  the   Galactic  disk,
including stars and  the interstellar medium. We set  the scale length
$l=3$\,kpc,  the  scale  height  $h=300$\,pc, and  its  local  surface
density  $\Sigma(R_0)=53\,\mathrm{M}_{\sun}  \mathrm{pc}^{-2}$ at  the
solar    radius   $R_0$,    which    includes   40$\,\mathrm{M}_{\sun}
\mathrm{pc}^{-2}$     for     the     stellar     contribution     and
13$\,\mathrm{M}_{\sun}  \mathrm{pc}^{-2}$ for  the ISM.   We neglected
the contributions of the bulge and the stellar halo, as these are very
small  beyond 3\,kpc  from the  Galactic centre.   To maintain  a flat
Galactic rotation  curve, we added  a Miyamoto spheroid  (Miyamoto and
Nagai,  \cite{miy75}).  Adopting  $R_0=8.5$\,kpc and  a  flat Galactic
rotation  curve $V_c(R=5\,to\,20\,$kpc$)=220\mathrm{\,km\,s}^{-1}$, we
adjusted the core radius ($a+b$) and the mass of the Miyamoto spheroid
to  find  $a+b=9.34$\,kpc.  If  the  Miyamoto  component is  spherical
($a$=0\,kpc),          its          local          density          is
$\rho_{\rm{d.m.}}=0.012\,\mathrm{M}_{\sun}           \mathrm{pc}^{-3}$.
Flattening  this dark component  ($a$=5\,kpc, i.e.   an axis  ratio of
0.51),    we    obtained    $\rho_{\rm{d.m.}}=0.021\,\mathrm{M}_{\sun}
\mathrm{pc}^{-3}$: the exact limit  compatible with our $K_z$ analysis
that  still  gives  good  fits  to  $\Sigma_{\rm  1.1\,kpc}$.   Larger
flattenings  are excluded,  as, for  instance, $a$=\,6.5\,kpc  gives a
0.34   axis   ratio   and   $\rho_{\rm{d.m.}}=0.030\,\mathrm{M}_{\sun}
\mathrm{pc}^{-3}$.

  In conclusion, our  data shows that the Galactic  dark matter can be
distributed  in a  spherical  component, but  it  certainly cannot  be
distributed in a  very flattened disk.  For instance,  if the Galaxy's
mass distribution  were totally flat, the local  surface density would
be   as  high   as  $211\,\mathrm{M}_{\sun}   \mathrm{pc}^{-2}$  (with
$V_c(R)=220$\,km\,s$^{-1}$ and  $R_0$=8.5\,kpc for a  Mestel disk, see
Binney  \& Tremaine  \cite{bin87}).  In  conclusion, there  is  room to
flatten  the dark  matter halo  by a  maximum factor  of about  two or
three.  This  agrees with the shape of the  dark halo that de
Boer (\cite{boer05}) interprets from diffuse Galactic EGRET gamma ray
excess for energies above 1\,GeV.

\subsection{Terrestrial  impact cratering}
\label{a:crater}

The main  topic of this paper  is the determination of  the local {\it
surface} mass density  $\Sigma(z)$: for that purpose we  used the most
recent  determinations   of  the  local  {\it   volume}  mass  density
$\rho_{\rm       total}(z=0)\,=\,0.10       \pm0.01\,\mathrm{M}_{\sun}
\mathrm{pc}^{-3}$.   Our data alone  gives a  less accurate  but close
value  of  $\rho_{\rm  total}(z=0)$\,=\,$0.1\pm0.02\,\mathrm{M}_{\sun}
\mathrm{pc}^{-3}$.  All these recent  determinations are based on very
different samples of stars: A-F  dwarfs and different types of giants.
These samples cover  different distances, either in a  sphere of 125\,pc
around  the Sun,  inside a  cylinder  of 800\,pc  length crossing  the
Galactic plane, or in pencil beams up to 1.1\,kpc from the plane.  All
these  determinations converge  towards the  same value  of  the local
volume   mass  density,   0.10$\,\mathrm{M}_{\sun}  \mathrm{pc}^{-3}$,
implying that the half period  of the vertical oscillations of the Sun
through the Galactic plane is 42$\pm2$\,Myr.

 A 26\,Myr periodicity in epochs  of major mass exctinction of species
was found by Raup \& Sepkoski (\cite{rau86}), and a cycle of 28.4\,Myr
in the  ages of  terrestrial impact craters  was found by  Alavarez \&
Muller  (\cite{alv84}).   The  periodicity  in these  catastrophes  is
disputed, however (see for instance Jetsu \& Pelt, \cite{jet00}).  The
spectral analysis  of the periodicity hypothesis  in cratering records
shows, in  the most recently published works,  possible or significant
periods:  $33\pm4$\,Myr  (Rampino  \& Stothers,  1984),  $33\pm1$\,Myr
(Stothers, \cite{sto98}), and more recently 16.1\,Myr and 34.7\,Myr by
Moon et al.  (\cite{moo03}).  We  note that some authors estimate that
periodicities could  result from  a spurious ``human-signal''  such as
rounding   (Jetsu  \&   Pelt,   \cite{jet00}).   Recently,   Yabushita
(\cite{yab02,yab04}) claims a periodicity of $37.5$\,Myr and considers
that the  probability of  deriving this period  by chance on  the null
hypothesis of  a random  distribution of crater  ages is  smaller than
0.10.  From the width of the  peaks in their periodograms, we estimate
their period accuracy to be $\pm2$\,Myr.

It  has  been  frequently  proposed  that the  period  of  high-impact
  terrestrial cratering  would be directly  linked to the  crossing of
  the  solar  system  through  the  Galactic  plane  where  the  giant
  molecular  clouds are  concentrated.   However, the  above-mentioned
  periods in the range 33-37\,Myr  would correspond to the half period
  of oscillation of the Sun only  if the total local mass density were
  0.15$\,\mathrm{M}_{\sun}  \mathrm{pc}^{-3}$,  a  high value  of  the
  local density measured  in the 1980's by a  few authors.  Recent and
  accurate   $\rho_{\rm   total}(z=0)$   measurements  now   imply   a
  42$\pm2\,$Myr  half period  of oscillation  of the  Sun that  can no
  longer be  related to possible  periods of large impact  craters and
  mass extinction events.\\

\section{Summary}

$\bullet$  Adopting  the previous  determination  of  the total  local
   volume  density  $\rho_{\rm total}(z=0)\,=\,0.10\,\mathrm{M}_{\sun}
   \mathrm{pc}^{-3}$ in the Galactic  disk, we modelled the vertical disk
   potential and mass distribution through the constraints of a sample
   of  red  clump  stars  towards  the NGP  with  measured  distances,
   velocities, and [Fe/H] abundances.

\noindent
$\bullet$ Our simplest model, including a spherical dark corona, shows
   that there is no need for extra dark matter in the Galactic disk to
   explain  the vertical gravitational  potential.  The  total surface
   mass   density,   at  the   Solar   position,   is   found  to   be
   $\Sigma_{1.1\,{\rm                 kpc}}=64\pm5\,\,\mathrm{M}_{\sun}
   \mathrm{pc}^{-2}$,   compared  to  53\,$\,\mathrm{M}_{\sun}
   \mathrm{pc}^{-2}$  for  the     visible  matter  (see  section
   \ref{a:stellarSM}), which is  sufficient to  explain  both the  observed
   amount of visible matter and the local contribution of a round dark
   matter halo (15\,$\,\mathrm{M}_{\sun} \mathrm{pc}^{-2}$).

\noindent
$\bullet$ With a  two-parameter model and some flattening  of the dark
   matter  halo, we obtained  a larger  extent of  acceptable solutions:
   $\Sigma_{1.1\,{\rm              kpc}}$\,=\,57-79\,$\mathrm{M}_{\sun}
   \mathrm{pc}^{-2}$,               $               \Sigma_{0.8\,{\rm
   kpc}}$\,=\,57-66\,$\mathrm{M}_{\sun}             \mathrm{pc}^{-2}$.
   Flattening larger than about 2 or  3 is excluded by  analysis of
   our red clump giants.

    We note that a flattening by a factor two of the dark corona could
     contradict the other recent  constraint obtained by Ibata
     et al.   (\cite{iba01}), which is based on the non-precessing  orbit of the
     Sagittarius  stream.  However,  while their  result  concerns the
     outer halo, our analysis is only sensitive to the inner one.

On the other hand, it is  not possible to flatten the dark matter halo
without increasing its own contribution to  the local volume
mass  density too  much.  The  recent  dynamical estimates  based on  Hipparcos
data    of   the    {\bf   total}    volume   mass    density   gives
0.10\,$\mathrm{M}_{\sun}  \mathrm{pc}^{-3}$. This  includes  the known
stellar     local      mass     density,     0.046\,$\mathrm{M}_{\sun}
\mathrm{pc}^{-3}$,    and    the     gas    volume    mass    density,
0.04\,$\mathrm{M}_{\sun}  \mathrm{pc}^{-3}$ (see  Palasi \cite{pal98},
Holmberg \&  Flynn \cite{hol00}, Chabrier  \cite{cha01,cha02,cha03}, or
Kaberla     \cite{kal03}).     This     leaves    room     for    only
0.014\,$\mathrm{M}_{\sun} \mathrm{pc}^{-3}$ for the dark matter.  This
also  implies that the  halo cannot  be flattened  more than  a factor
$\sim$two,  unless the  volume mass  density of  the gas,  the weakest
point in the $K_z$ analysis, has been strongly overestimated.

\noindent
$\bullet$ As a by-product of this  study we  determined  the half
  period  of  oscillation  of  the  Sun through  the  Galactic  plane,
  $42\pm2\,$Myr,  which cannot be  related to  the possible  period of
  large terrestrial impact craters $\sim$ 33-37\,Myr.

\begin{acknowledgements}
This  research  has made  use  of  the  SIMBAD and  VIZIER  databases,
operated at the CDS, Strasbourg, France.  It is based on data from the
ESA  {\it  Hipparcos} satellite  (Hipparcos  and Tycho-2  catalogues).
Special thanks  go to P.  Girard,  C.  Boily, and L.   Veltz for their
participation in  the observations  and to A.  Robin and C.  Flynn for
useful comments.

\end{acknowledgements}


\begin{thebibliography}{}

\bibitem[1984]{alv84}
Alvarez, W., Muller, R.A., 1984, Nature, 308, 718

\bibitem[1999]{are99}
Arenou, F.,  Luri, X. 1999, in Harmonizing Cosmic Distance Scales in a Post-Hipparcos Era, ASP Conf. Ser., 167, 13

\bibitem[1984]{bah84}
Bahcall, J. N. 1984, ApJ, 276, 156

\bibitem[2000]{BB00}
{Barbier-Brossat}, M. \& {Figon}, P. 2000, A\&A Sup., 142, 217

\bibitem[1999]{bie99}
Bienaym\'e, O. 1999, A\&A, 341, 86

\bibitem[1987]{bie87}
Bienaym\'e, O., Robin, A., Cr\'ez\'e, M. 1987 A\&A, 186, 359

\bibitem[1987]{bin87}
Binney, J., Tremaine, S. 1987, Galactic Dynamics, Princeton Series in Astrophysics

\bibitem[2005]{boer05}
de Boer, W. 2005, New Astr. Rev., 49, 213

\bibitem[1996]{CFFST96}
 Cayrel, R., Faurobert-Sholl, M., Feautrier, N., Spielfield, A., Thevenin, F.
 1996, A\&A 312, 549

\bibitem[2001]{cayrelstr01}
 Cayrel de Strobel, G., Soubiran, C., Ralite, N. 2001, A\&A, 373, 159

\bibitem[2001]{cha01}
Chabrier, G. 2001, ApJ, 554, 1274

\bibitem[2002]{cha02}
Chabrier, G. 2002, ApJ, 567, 304 

\bibitem[2003]{cha03} 
Chabrier, G. 2002, ApJ, 586, L133

\bibitem[2003]{che03} 
Chen, A. B.-C., Lu, P.K., M\'endez, R. A., van Altena, W.F. 2003, AJ, 126, 762

\bibitem[1999]{che99}
Chereul, E., Cr\'ez\'e, M., Bienaym\'e, O. 1999, A\&A, 135, 5

\bibitem[1998a]{cre98a}
Cr\'ez\'e, M., Chereul, E., Bienaym\'e, O., \& Pichon, C. 1998a, A\&A, 329, 920

\bibitem[1998b]{cre98b}
Cr\'ez\'e, M., Chereul, E., Bienaym\'e, O., \& Pichon, C. 1998b, Proc. of the ESA Symp. "Hipparcos - Venice 97'', ESA SP-402, 669

\bibitem[ESA, 1997]{esa97}
ESA 1997, The Hipparcos and Tycho Catalogues, (Noordwijk) Series: ESA-SP 1200

\bibitem[2004]{fam04}
Famaey, B., Jorissen, A., Luri, X. et al. 2004, A\&A, 430, 165

\bibitem[1994]{fly94}
Flynn, C.,  Fuchs, B. 1994, MNRAS, 270, 471

\bibitem[1994]{gal94}
 Galazutdinov,  A.G.,  1994, Odessa  Astron.  Pub.,  7, 88

\bibitem[1990]{gou90}
Gould, A. 1990, ApJ, 360, 504

\bibitem[1997]{gou97}
Gould, A., Bahcall, J., Flynn, C. 1997, ApJ, 482, 913

\bibitem[1994]{gray94}
Gray, D. 1994, PASP 106, 1248

\bibitem[1998]{har98} 
Harmanec, P., 1998,  A\&A, 335, 173

\bibitem[2001]{hay01} 
Haywood, M., 2001, MNRAS, 325, 1365

\bibitem[2002]{hay02} 
Haywood, M., 2002, MNRAS, 337, 151

\bibitem[2000]{hog00} 
H{\o}g, E., Fabricius, C., Makarov, V.\,V. et al. 2000, A\&A, 363, 385

\bibitem[1998]{hog98}
H{\o}g, E., Flynn, C. 1998, MNRAS, 294, 28

\bibitem[2000]{hol00}
Holmberg, J., Flynn, C. 2000, MNRAS, 313, 209

\bibitem[2004]{hol04}
Holmberg, J., Flynn, C. 2004, MNRAS, 352, 440

\bibitem[2001]{iba01} 
Ibata, R.,  Lewis, G.F., Irwin, M. et al. 2001, ApJ, 551, 294

\bibitem[2000]{jet00}
Jetsu, L., Pelt, J. 2000, A\&A, 353, 409

\bibitem[2003]{kal03} 
Kalberla, P.M.W. 2003, ApJ, 588, 823

\bibitem[1922]{kap22}
Kapteyn, J. C. 1922, ApJ, 55, 302

\bibitem[1998]{kat98} 
Katz, D., Soubiran, C., Cayrel, R. et al.  1998 A\&A, 338, 151

\bibitem[2003]{kor03}
Korchagin, V.\,I., Girard, T.\,M., Borkova, T.\,V., Dinescu, D.\,I., van Altena, W.\,F. 2003, AJ, 126, 2896

\bibitem[2005]{Kovtet05}
Kovtyukh, V.V., Mishenina, T.V., Gorbaneva, T.I., Bienaym\'e O., Soubiran, C.,
Kantcen, L.E. 2005, Sov. Astron. (in press)

\bibitem[1991]{kui91} 
Kuijken, K. 1991, ApJ, 372, 125

\bibitem[1989]{kg89}
Kuijken, K., Gilmore, G. 1989, MNRAS, 239, 605

\bibitem[2005]{Mishet05}
Mishenina, T.V. et al. 2005 in preparation

\bibitem[1975]{miy75}
Miyamoto, M. Nagai, R. 1975, PASJ, 27, 533

\bibitem[1990]{mcW90}
McWilliam, A. 1990, ApJS 74, 1075

\bibitem[2003]{moo03}
Moon, H.K., Min, B.H., Kim, S.L. 2003, J Astron. Space Sci., 20(4), 269

\bibitem[1932]{oor32}
Oort, J. H. 1932, BAN, 6, 249

\bibitem[1960]{oor60}
Oort, J. H. 1960, BAN, 15, 45

\bibitem[1998]{pal98} 
Palasi, J. 1998, Proc. of the ESA Symp. "Hipparcos - Venice 97'', ESA SP-402, 551

\bibitem[2001]{per01}
Perryman, M.A.C., de Boer, K.S., Gilmore, G. et al. 2001 A\&A 369, 339

\bibitem[1998]{pha98}
Pham, H.-A. 1998, Proc. of the ESA Symp. "Hipparcos - Venice 97'', ESA SP-402, 559

\bibitem[1989]{phi89} 
Philip  A.G.D.,  Lu,  P.K. (Eds.),  1989  The Gravitational  Force
Perpendicular  to   the  Galactic   Plane  (1989),  L.   Davis  Press,
Schenectady, N.Y.

\bibitem[2001]{pru01} 
Prugniel, Ph., Soubiran, C. 2001,  A\&A, 369, 1048

\bibitem[1984]{ram84} 
Rampino, M.R., Stothers, R.B., 1984, Nature, 308, 709

\bibitem[1986]{rau86} 
Raup, D.M., Sepkoski, J.J.  1986, Sci, 231, 833

\bibitem[2003]{red03} 
Reddy, B.E., Tomkin, J., Lambert, D. L., Allende Prieto, C. 2003, MNRAS, 340, 304

\bibitem[2003]{rob03} 
Robin, A. C., Reyl\'e, C., Derri\`ere, S., Picaud, S.
2003, A\&A, 409, 523

\bibitem[2004]{set04} 
Setiawan, J., Pasquini, L., da Silva et al. 2004, A\&A, 421, 241

\bibitem[2003]{sie03}
Siebert, A., Bienaym\'e, O., Soubiran, C. 2003, A\&A, 399, 531 (paper II)

\bibitem[2004]{smi04}
Smith, B., Price, S., Baker, R.. 2004, ApJS, 154, 673
 
  \bibitem[2003]{sou03} 
Soubiran, C., Bienaym\'e, O., Siebert, A. 2003, A\&A, 398, 141 (paper I)

  \bibitem[1998]{sou98} 
Soubiran, C., Katz, D., Cayrel, R. 1998, A\&A, 133, 221

 \bibitem[2005]{sou05} 
Soubiran, C. et al. 2005 in preparation  (paper IV)

\bibitem[2003]{ste03} 
Steinmetz,  M. 2003, GAIA  Spectroscopy: Science
and Technology, ASP\,Conf.\,Proc., Vol. 298, Ed. U.\,Munari, 381

\bibitem[1998]{sto98}
Stothers, R. 1998, MNRAS, 300, 1098

\bibitem[1996]{tsymbal96}
Tsymbal, V.V.  1996, Model Atmospheres and Spectrum Synthesis,
 ASP Conf. Ser. 108, 198

\bibitem[2001]{zhaoet01}
Zhao, G., Qiu, H.M., Mao, S. 2001, ApJ, 551, L85

\bibitem[2002]{yab02}
Yabushita, S. 2002, MNRAS, 334, 369

\bibitem[2004]{yab04}
Yabushita, S. 2004, MNRAS, 355, 51

\bibitem[2001]{zhe01}
Zheng, Z., Flynn, C., Gould, A. et al. 2001, ApJ, 555, 393

\end{thebibliography}
\end{document}